\documentclass[prl, reprint, amsmath,amssymb,nofootinbib,floatfix, aps,longbibliography,  superscriptaddress, preprintnumbers]{revtex4-1}
\usepackage{graphicx}
\usepackage{dcolumn}
\usepackage{bbold}
\usepackage{mathtools}
\usepackage[colorlinks=true,urlcolor=rossos,anchorcolor=black,citecolor=mygreen,filecolor=black,linkcolor=black,menucolor=blue,linktocpage=true]{hyperref} 
\usepackage{booktabs}
\usepackage{colortbl}
\usepackage{collcell}
\usepackage{enumerate}
\usepackage{float}
\usepackage{verbatim}
\usepackage{multirow}
\usepackage{xparse}
\usepackage{slashed}
\usepackage{tabularx}

\definecolor{rossos}{cmyk}{0,1,1,0.55}
\definecolor{mygreen}{rgb}{0.27, 0.64, 0.48}
\definecolor{mygray}{gray}{.95}

\newcommand\sh{\sinh}

\newcommand{\virg}[1]{``#1''}

\begin{document}

\title{Low-Scale Leptogenesis with Low-Energy Dirac CP-Violation 
}

 \author{Alessandro Granelli}
 \email{alessandro.granelli@unibo.it}
 \affiliation{Dipartimento di Fisica e Astronomia, Università di Bologna, via Irnerio 46, 40126, Bologna, Italy}
 \affiliation{INFN, Sezione di Bologna, viale Berti Pichat 6/2, 40127, Bologna, Italy}
\author{Silvia Pascoli}
 \affiliation{Dipartimento di Fisica e Astronomia, Università di Bologna, via Irnerio 46, 40126, Bologna, Italy}
 \affiliation{INFN, Sezione di Bologna, viale Berti Pichat 6/2, 40127, 
Bologna, Italy}
 \author{Serguey T.~Petcov}
  \altaffiliation{Also at: Institute of Nuclear Research and Nuclear Energy, Bulgarian Academy of Sciences, 1784 Sofia, Bulgaria.}
 \affiliation{INFN/SISSA, via Bonomea 265, 34136, Trieste, Italy}
 \affiliation{Kavli IPMU (WPI), UTIAS, University of Tokyo, Kashiwa,
  Chiba 277-8583, Japan.}

 \begin{abstract}
We study the freeze-in  scenario of leptogenesis via oscillations within the type-I seesaw model with two quasi-degenerate heavy Majorana neutrinos $N_{1,\,2}$ having masses $M_2 > M_1 \sim (0.1-100)\,\text{GeV}$, $(M_2-M_1)/M_1 \ll 1$, focusing on the role of the CP-violation provided by the Dirac phase $\delta$ of the Pontecorvo-Maki-Nakagawa-Sakata lepton 
mixing matrix. 
We find that viable leptogenesis can be due solely to CP-violating values of $\delta$ and that the $N_{1,\,2}$ total mixing squared $\Theta^2=\sum_\alpha\Theta^2_\alpha$ needed is within the reach of future experiments, $\Theta_\alpha$ parameterising the coupling to the charged lepton $\alpha=e,\,\mu,\,\tau$. Furthermore, the required parameter space differs from that associated with additional Casas-Ibarra sources of CP-violation.
Future determination of $\delta$, $\Theta^2$ and/or the ratios $\Theta_\tau^2:\Theta^2_\mu:\Theta^2_e$ would provide a critical test of the considered scenario.

 \end{abstract}
\maketitle

\noindent \textit{\textbf{Introduction}}--- 
In the present observable Universe there is an overabundance of matter over antimatter. The asymmetry in baryons, or the \textit{baryon asymmetry of the Universe} (BAU), can be parameterised by the baryon-to-photon ratio $\eta_B$.
Observations of the cosmic microwave background anisotropies and the abundances of light primordial elements agree on the present value of $\eta_B\simeq 6.1\times 10^{-10}$ \cite{Planck2018, Cooke_2018}. An early mechanism to generate the BAU is referred to as \textit{baryogenesis} (see \cite{Bodeker:2020ghk} for a recent review), but it is unfeasible within the \textit{Standard Model} (SM) of particle physics and new physics is required.

An alternative attractive mechanism is that of baryogenesis via
\textit{leptogenesis} (LG) \cite{Fukugita:1986hr}, consisting of an early generation of a lepton asymmetry, which is then converted into the present BAU by the SM sphaleron processes \cite{Kuzmin:1985mm}. The simplest scenario of LG is realised within the type-I seesaw extension of the SM \cite{Minkowski:1977sc,Yanagida:1979as,GellMann:1980vs,Glashow:1979nm,Mohapatra:1979ia}, which also provides a mechanism for the generation of the light neutrino masses by augmenting the SM with right-handed sterile neutrinos. 
The type-I seesaw extension with two right-handed neutrinos
and, correspondingly, with two heavy Majorana neutrinos $N_{1,\,2}$ with definite masses $M_{1,\,2} > 0$, is the minimal set-up in which LG can be realised, while being also compatible with current data on light neutrino masses and mixing. 

Many realisations of LG within the type-I seesaw extension are possible depending on the mass scale \cite{Fukugita:1986hr, Kuzmin:1985mm, Pilaftsis:1997jf, Pilaftsis:2003gt, Akhmedov:1998qx, Asaka:2005pn, Racker:2012vw}, through lepton number, C- and CP-violating, out-of-equilibrium processes involving 
the heavy Majorana neutrinos, the Higgs and left-handed lepton doublets, 
which satisfy the necessary Sakharov's conditions \cite{Sakharov:1967dj}. In this work, we are 
focused on the
 \virg{freeze-in} mechanism proposed in \cite{Akhmedov:1998qx, Asaka:2005pn} and extensively studied \cite{hepph0605047,Asaka:2011wq, canetti2013dark,Shuve:2014zua,1508.03676,Drewes:2016gmt, Hernandez:2016kel, Drewes:2016jae,1704.02692,Ghiglieri:2017gjz,Drewes:2017zyw,Abada:2018oly, Klaric:2020phc, Klaric:2021cpi, Hernandez:2022ivz, Drewes:2022akb, Sandner:2023tcg}, in which the oscillations of the right-handed neutrinos during their out-of-equilibrium production are crucial for the generation of the BAU. 
This scenario of \textit{LG via oscillations} 
can be successful for heavy Majorana neutrinos mass scales
as low as $100\,\text{MeV}$, thus being accessible to low-energy searches of heavy neutral leptons \cite{Abdullahi:2022jlv, Antel:2023hkf}.

Among the type-I seesaw model parameters, there are multiple CP-violating phases that could provide the CP-violation necessary for successful LG. Up to a change of basis, the CP-violating phases can be recast inside the matrix $Y$ of the Yukawa coupling between $N_{1,\,2}$, the Higgs and the left-handed lepton doublets. Under the widely-adopted \textit{Casas-Ibarra parameterisation} \cite{Casas:2001sr}, the Yukawa matrix can be written as 
\begin{equation}\label{eq:CI_matrix}
    Y = \pm i \frac{\sqrt{2}}{v} U\sqrt{{\hat{m}}_\nu} O^T \sqrt{{\hat{M}}_N}\,,
\end{equation} 
where $U$ is the Pontecorvo-Maki-Nakagawa-Sakata (PMNS) lepton mixing matrix, $\hat{m}_\nu \equiv \text{diag}(m_1,m_2,m_3)$, with $0\leq m_{1,\,2,\,3} \ll M_{1,\,2}$, is the diagonal mass matrix for the light neutrinos, and $\hat{M}_N = \text{diag}(M_1, M_2)$. The complex \textit{Casas-Ibarra matrix} $O$, in the case of two heavy Majorana neutrinos, is $2\times3$ with orthonormal rows and it can be parameterised in terms of an arbitrary complex angle, thus containing CP-violating phases. With $Y$ written as in Eq.~\eqref{eq:CI_matrix}, there is an explicit distinction between the CP-violating Dirac and Majorana phases of the PMNS matrix, which are associated to low-energy phenomenology, and those of the Casas-Ibarra matrix, which can manifest themselves in physical processes involving the heavy Majorana neutrinos.

A physically interesting possibility is when the requisite 
CP-violation in LG is only due to the phases of the PMNS
lepton mixing matrix \cite{Pascoli:2006ie, Pascoli:2006ci, Blanchet:2006be, Branco:2006ce, Uhlig:2006xf, Anisimov:2007mw, Molinaro:2009lud, Molinaro:2008cw, Bambhaniya:2016rbb, Dolan:2018qpy, Moffat:2018smo, Granelli:2021fyc}. In this case, there would be a direct link between the BAU and CP-violating phenomena in low-energy neutrino physics, such as, e.g., in neutrino oscillations or in neutrinoless double beta
decay (see e.g., \cite{Tanabashi:2018oca}).
At present, only indications of CP-violation in neutrino oscillations involving the Dirac phase $\delta$ exist. However, $\delta$ is determined in the global analyses with relatively large uncertainties \cite{Capozzi_2020, Esteban:2020cvm, nufit} and CP-conserving values are not yet excluded. Current experiments such as T2K \cite{T2K:2011qtm} and NO$\nu$A \cite{NOvA:2021nfi} will be able to provide additional information in the next future, potentially reaching $\sim 3\sigma$ for hints of CP-violation.
The experiments DUNE \cite{DUNE:2021tad}, and T2Hyper-Kamiokande (T2HK) \cite{Hyper-Kamiokande:2022smq}, currently under construction, will have much stronger sensitivity, aiming at a 5$\sigma$ discovery of leptonic CP-violation for a large fraction of the possible values of $\delta$.

Within the type-I seesaw extension, the phases of the PMNS matrix are the unique sources of CP-violation in the neutrino sector when the Casas-Ibarra matrix is CP-conserving \cite{Pascoli:2006ie, Pascoli:2006ci}. This condition corresponds to specific classes of seesaw models in which the elements of the matrix $O$ are either real or purely imaginary and it can be realised, for instance, in flavour models based on sequential dominance \cite{King:2006hn} or with residual CP-symmetries \cite{Chen:2016ptr, Hagedorn:2016lva}.

It is possible that the Dirac phase $\delta$ is the only source 
of CP-violation in the lepton sector. 
LG with Dirac CP-violation has been shown to work in the thermal high-scale scenarios \cite{Pascoli:2006ci, Anisimov:2007mw, Molinaro:2009lud, Molinaro:2008cw, Dolan:2018qpy, Moffat:2018smo, Granelli:2021fyc}, emerging as one of the motivations for the current and future neutrino oscillation experimental programme. As great attention is being put to the searches of heavy neutral leptons at the GeV scale \cite{Abdullahi:2022jlv, Antel:2023hkf}, the question on whether low-scale LG via oscillations can be successful with low-energy CP-violation solely from the Dirac phase should be answered also in this context. In this paper, we examine this 
physically interesting possibility with particular attention 
to the related low-energy phenomenology. This could serve as further motivations for neutrino oscillation experiments and suggest new directions for heavy neutral lepton searches.\\

\noindent \textit{\textbf{The Framework}}--- We consider the 
minimal version of the type-I seesaw extension of the SM 
with two heavy Majorana neutrinos neutrinos $N_{1,2}$ having masses 
$M_2 > M_1\sim (0.1 - 100)$ GeV and a mass splitting 
$\Delta M\equiv M_2 - M_1\ll M_1$ 
in the range $\Delta M/M_1 \sim (10^{-11} - 10^{-4})$.
In the type-I seesaw, 
after the neutral component of the Higgs doublet acquires a non-vanishing 
vacuum expectation value $v = 246$ GeV, one gets
the well known relation 
$\left(m_\nu\right)_{\alpha \beta} 
\simeq -(v^2/2)\,\sum_{j=1,2}Y_{\alpha j}Y_{\beta j}~M_j^{-1}$, 
$\alpha,\beta =e,\,\mu,\,\tau$, for the entries of the tree-level 
light neutrino mass matrix $m_\nu$, where $Y_{\alpha j}$ is the Yukawa coupling 
of $N_j$ with the Higgs and left-handed lepton doublet of flavour $\alpha$.
The matrix $m_\nu$ can be diagonalised as $\hat{m}_\nu = U^\dagger m_\nu U^*$, 
where $\hat{m}_\nu \equiv \text{diag}(m_1,\,m_2,\,m_3)$ and $U$ represents 
the PMNS lepton
mixing matrix.
We adopt the standard parameterisation for $U$ \cite{Tanabashi:2018oca}
in terms of three neutrino mixing angles $\theta_{12}$, $\theta_{23}$ and 
$\theta_{13}$, the Dirac phase $\delta$, and two Majorana phases $\alpha_{21}$ 
and $\alpha_{31}$ \cite{Bilenky:1980cx}.
In the case of 
two heavy Majorana neutrinos, the lightest 
neutrino is massless at tree and 
one-loop levels and the light neutrino mass spectrum is hierarchical 
with either normal ordering (NO) $m_1\simeq 0 \ll m_2 < m_3$, or 
inverted ordering (IO)  $m_3\simeq 0 \ll m_1 < m_2$.
In the numerical analysis that follows, we consider the best-fit values 
of $\theta_{12}$, $\theta_{23}$ and $\theta_{13}$, and the two neutrino mass 
squared differences obtained 
in 
\cite{nufit, Esteban:2020cvm},
but treat 
$\delta$ as a free parameter due to the relatively large uncertainty in its determination. The Majorana phases $\alpha_{21}$ and 
$\alpha_{31}$ cannot be constrained by the neutrino oscillation 
experiments \cite{Bilenky:1980cx}
and are undetermined at present. In the studied case, 
only the combination $\alpha_{23}\equiv\alpha_{21}-\alpha_{31}$ 
(the phase $\alpha_{21})$ is physical in the hierarchical NO (IO) case. 
We treat $\alpha_{23(21)}$ as free parameters. For reasons that will be clearer throughout the text, we concentrate the analysis mostly on the light-neutrino mass spectrum with NO and leave the IO case for a future longer work. Global analyses including data from atmospheric, reactor and long-baseline neutrino experiments give a mild preference for NO against the spectrum with IO \cite{Capozzi_2020, Esteban:2020cvm}.

We consider the Casas-Ibarra (CI) parameterisation for the Yukawa matrix
\cite{Casas:2001sr}, which we rewrite explicitly as:
$Y_{\alpha j} = \pm i (\sqrt{2}/v) \sum_{a=1,\,2,\,3}U_{\alpha a}\sqrt{m_a}O_{ja}\sqrt{M_j}\,.
$
%
The arbitrary 
CI matrix $O$
have entries $O_{11(13)} = O_{21(23)} =0$, 
$O_{23(22)} = \varphi \,O_{12(11)} = \varphi \cos\theta$ 
and $O_{13(12)} = -\varphi \,O_{22(21)} = \varphi \sin\theta$ in the NO (IO) case,
with $\theta \equiv \omega + i \xi$, $\omega$ and $\xi$ being free real 
parameters and $\varphi = \pm1$. 
We choose to work with $\varphi = +1$ but extend the range of the Majorana phases $\alpha_{23(21)}$ from $[0, 2\pi]$ to $[0, 4\pi]$. In this way, the same full sets of CI and Yukawa matrices are considered \cite{Molinaro:2008rg}. 

The SM flavour neutrinos also mix with the heavy Majorana neutrinos. 
The mixing $\Theta_{\alpha j} \simeq (v/\sqrt{2})Y_{\alpha j}/M_j$ sets the 
coupling between $N_j$ and 
the charged lepton $\alpha$ ($\nu_\alpha$) in the weak charged 
(neutral) current,  
thus being important for low-energy phenomenology. 
For instance, direct searches at colliders, beam-dump and kaon experiments are sensitive to 
$\Theta^2_\alpha \equiv \sum_{j=1}^2 |\Theta_{\alpha j}|^2$ and 
$\Theta^2 \equiv \sum_{\alpha = e,\,\mu,\,\tau}\Theta_\alpha^2$. 
The same quantities are crucial in 
LG as they determine 
the strength of the \textit{wash-out} processes.\\

\noindent \textit{\textbf{Low-Energy CP-Violation}}--- Within the considered 
CI parameterisation, the CP-violating matrices can be either $U$, $O$ or both. 
In the case of \textit{low-energy CP-violation} (LECPV) we are interested in, 
the only CP-violating matrix is $U$, with the CI matrix being CP-conserving. 
LECPV can be achieved \cite{Pascoli:2006ci}
either by setting i) $\xi=0$ and $\omega\neq 0$, 
with real CI matrix; 
or ii) $\omega = k\pi$, $k = 0, 1/2, 1, ...$, 
and $\xi\neq 0$, so that $O_{12}O_{13}$ ($O_{11}O_{12}$) in the NO (IO) case is purely imaginary. 
Case ii) is associated with relatively large values of the mixings 
$\Theta^2_a$ and $\Theta^2$, as the condition $|\xi|\gg 1$ leads to an overall 
exponential enhancement. Since we are interested in connecting with 
experimental searches of heavy Majorana neutrinos, we focus the analysis on 
the case with $\omega = k\pi$ and $\xi\neq 0$. We stress that the
condition $\omega\neq k\pi$ when $\xi\neq 0$ 
would result in a CP-violating CI matrix (CICPV) \cite{Pascoli:2006ci}.

To have LECPV, the phases in the PMNS matrix  
should be CP-violating, i.e., 
$\delta \neq 0,\,\pi$, and/or $\alpha_{21} \neq k_{21}\pi$ and/or
$\alpha_{31} \neq k_{31}\pi$,
$k_{21} = 0,\,1,\,2,\,...$, $k_{31} = 0,\,1,\,2,\,...$. It is also possible, however, that CP is violated even when $U$ and $O$ 
are CP-conserving, but $Y$ is not \cite{Pascoli:2006ci}. 
In this case, CP is broken due to an 
\textit{interplay} between the PMNS and CI matrices in the CI parameterisation 
of the Yukawa matrix. When $\xi\neq 0 $ and $\omega = k\pi$, this can be realised 
for the CP-conserving values of the PMNS phases satisfying, additionally,
$\alpha_{23} \neq \pm (2n + 1)\pi$ ($\alpha_{21} \neq (2n + 1)\pi$), 
$n = 0,\,1$, in the NO (IO) case \cite{Pascoli:2006ci}. For the purpose of 
studying the case of LECPV uniquely from $\delta$, 
we shall consider $\alpha_{23(21)} = \pi$ or $3\pi$.\\

\noindent \textit{\textbf{CP-Violation in Leptogenesis}}--- All the 
CP-violating physical observables are expected to depend upon specific 
basis-independent quantities written in terms of the flavour parameters of 
the model, the so-called \textit{CP-violating invariants}.
For instance, the magnitude of CP-violation in
$\nu_\alpha \rightarrow \nu_{\beta}$ and
$\bar{\nu}_\alpha \rightarrow \bar{\nu}_{\beta}$
oscillations ($\alpha\neq \beta$)
is determined by the rephasing invariant $J_\text{CP} =
\Im\left[U_{\mu 3}\,U^*_{e3}\,U_{e2}\,U^*_{\mu 2}\right ]$ \cite{Krastev:1988yu},  
analogous to the Jarlskog invariant 
in the quark sector \cite{Jarlskog:1985ht, Jarlskog:1985cw, Bernabeu:1986fc}. 
Several
CP-invariants can be derived in the type-I seesaw extension of the SM 
starting from the Yukawa and heavy Majorana neutrino mass matrices 
\cite{Branco:2001pq, Branco:2004hu, Jenkins:2007ip, Jenkins:2009dy, Wang:2021wdq, Yu:2021cco}, and those that are relevant to LG (at leading order and in 
the case of two quasi-degenerate in mass 
heavy Majorana neutrinos) 
can be constructed out of the following two building blocks 
(see \cite{Hernandez:2022ivz} for a recent derivation):
$ J^\text{LNC}_\alpha = \Im\left[Y_{\alpha 1}^*Y_{\alpha 2}(Y^\dagger Y)_{21}\right]$
and
$J^\text{LNV}_\alpha = \Im\left[Y_{\alpha 1}^*Y_{\alpha 2}(Y^\dagger Y)_{12}\right]$.
At leading order, the BAU arising in LG is proportional to a combination of $J^\text{LNC}_\alpha$ and $J_\alpha^\text{LNV}$ weighted over the lepton flavours \cite{Flanz:1994yx, Covi:1996fm, Buchmuller:1997yu, Hambye:2017elz, Klaric:2021cpi, Hernandez:2022ivz}. For LECPV with $\omega = k\pi$ and $\xi\neq 0$, 
we have that $J^\text{LNC}_\alpha =  -J^\text{LNV}_\alpha \propto \Re[U_{\alpha 3(2)}^*U_{\alpha 2(1)}]\sh(2\xi)\cos(2k\pi)$ in the NO (IO) case.

For $M_1\gtrsim 100\,\text{GeV}$, outside the mass range of interest to this study,
low-scale LG has been shown to reconnect with the resonant freeze-out 
mechanism \cite{Klaric:2020phc, Klaric:2021cpi}. 
In the resonant LG scenario and within the Boltzmann equations formalism, the lepton asymmetry of flavour 
$\alpha$ is proportional to the sum of the two invariants 
\cite{Flanz:1994yx, COVI1996169,Covi:1996fm, Granelli:2020ysj, Klaric:2021cpi} 
$J_\alpha^\text{LNC} + J_\alpha^\text{LNV}\propto \sin(2\omega)$ up to corrections 
of the order of $\mathcal{O}(\Delta M/M_1)$, which vanishes when 
$\omega = k\pi$ contrarily to what happens in the low-scale LG scenario via 
oscillations. This highlights the importance of the oscillation mechanism 
in the considered framework.

We further note that, in the IO case, 
$\Re[U_{e 2}^*U_{e 1}]\propto \cos(\alpha_{21}/2)$, so that, when 
$\alpha_{21} = \pi,\,3\pi$, $J_e^\text{LNC} = J_e^\text{LNV} = 0$, 
$J_\mu^\text{LNC} = -J_\tau^\text{LNC}$ and $J_\mu^\text{LNV} = -J_\tau^\text{LNV}$. 
In this case, higher order CP-invariants can be relevant to LG, making the IO case more involved. \\

%
\noindent \textit{\textbf{Results}}--- We perform a numerical scan of 
the parameter space of viable LG. To calculate the BAU in the scenario 
of interest, we solve the momentum-averaged \textit{density matrix equations} 
\cite{Akhmedov:1998qx,Asaka:2005pn,Asaka:2011wq,Canetti:2012kh,Hernandez:2016kel,Hambye:2017elz,Ghiglieri:2017gjz,Ghiglieri:2017csp,Eijima:2018qke,Abada:2018oly,Klaric:2020phc,Klaric:2021cpi} for the evolution of the lepton asymmetries 
and heavy Majorana neutrino abundances. We consider the equations as 
in \cite{Abada:2018oly, Hernandez:2022ivz, Sandner:2023tcg} and make use of 
the latest version of the \texttt{ULYSSES} Python package 
\cite{Granelli:2020pim, Granelli:2023vcm}. We list in what follows the results of our numerical analysis.

\begin{itemize}
\item{We show in Fig.~\ref{fig:ScanLG_NH_dirac} the region in 
the $\Theta^2-M_1$ plane 
where LG with LECPV from $\delta$ 
is successful in reproducing
the observed value 
of the BAU. For illustrative purposes, we choose $\delta = 3\pi/2$, 
$\alpha_{23}= \pi$ and $\omega = 0$, vary $\Delta M/M_1$ in the range 
$[10^{-11},\,10^{-4}]$ and focus on the NO case. 
A qualitatively similar figure for the same choice of parameters 
is obtained in the IO case, but not shown here. 

The upper (lower) solid black line in the plot is the curve of maximal 
(minimal) mixing $\Theta^2$ compatible with viable LG. 
The shaded blue area between the two black lines correspond to successful LG 
for certain choices of $\Delta M/M_1$ and $\delta$. We paint in darker 
(lighter) blue the regions of successful LG corresponding to larger (smaller) 
values of $\Delta M /M_1$. 
We find that the extreme values of $\Theta^2$ can be obtained for 
$\Delta M/M_1\lesssim 10^{-6}$, while, for larger splittings, the viable 
region reduces in size, with the maximal (minimal) allowed mixing 
taking smaller (larger) values. 
%
%
    \begin{figure}
        \centering
        \includegraphics[width = 0.476\textwidth]{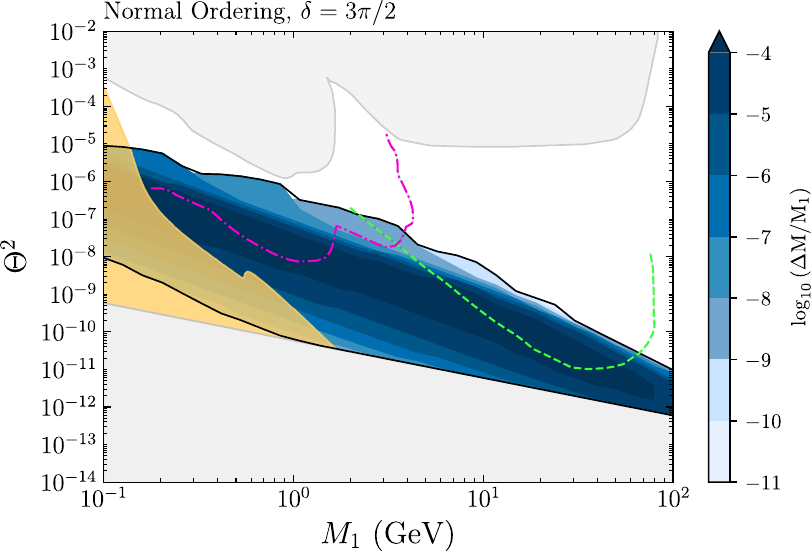}
        \caption{The parameter space of viable LG with LECPV solely from 
        $\delta$, in the NO case, for $\omega = 0$, $\alpha_{23} = \pi$ and $\delta = 3\pi/2$. The lower grey area is forbidden in the type-I seesaw mechanism of light neutrino mass generation. The upper grey region is a combination of current constraints on $\Theta^2_\tau$ \cite{CHARM:1985anb, DELPHI:1996qcc, T2K:2019jwa, ArgoNeuT:2021clc, BaBar:2022cqj, Barouki:2022bkt}, the yellow one is excluded by BBN \cite{Sabti:2020yrt, Boyarsky:2020dzc}. The dot-dashed purple line represents the expected sensitivities of several upcoming and proposed experiments \cite{FASER:2018eoc, Dib:2019tuj, Aielli:2019ivi, Batell:2020vqn, HIKE:2022qra, Abdullahi:2022jlv, Aberle:2839677, MATHUSLA:2020uve, Antel:2023hkf}, while the green dashed one is  
that of FCC-ee \cite{FCC:2018byv, FCC:2018evy}.}
        \label{fig:ScanLG_NH_dirac}
    \end{figure}

The LG parameter space is bounded from below by the requirement of reproducing 
the light neutrino masses (lower grey region) and from above by the 
experimental  
limits on the couplings of the heavy Majorana neutrinos to
the electron 
\cite{CHARM:1985anb, Bernardi:1987ek, DELPHI:1996qcc, Belle:2013ytx, PIENU:2017wbj, CMS:2018iaf, T2K:2019jwa, ATLAS:2019kpx, NA62:2020xlg, CMS:2022fut, ATLAS:2022atq}, the muon \cite{CHARM:1985anb, Bernardi:1987ek, DELPHI:1996qcc, Belle:2013ytx, PIENU:2017wbj, CMS:2018iaf, T2K:2019jwa, ATLAS:2019kpx, MicroBooNE:2019izn, NA62:2021bji, CMS:2022fut, ATLAS:2022atq, MicroBooNE:2022ctm} and 
the tauon \cite{CHARM:1985anb, DELPHI:1996qcc, T2K:2019jwa, ArgoNeuT:2021clc, BaBar:2022cqj, Barouki:2022bkt} flavour. Numerous planned and proposed 
experiments aim at improving the sensitivity to 
these couplings
further \cite{FASER:2018eoc, Beacham:2019nyx, Dib:2019tuj, Aielli:2019ivi, Batell:2020vqn, MATHUSLA:2020uve, PIONEER:2022yag, HIKE:2022qra, Blinov:2021say, Abdullahi:2022jlv, Aberle:2839677, Alviggi:2839484}. 
The total mixing is also constrained by the Big Bang Nucleosynthesis (BBN) 
\cite{Sabti:2020yrt, Boyarsky:2020dzc}. The reported limits and projections on 
$\Theta^2$, however, are currently based on the assumption that only 
the mixing in a particular flavour $\alpha$ is non-zero, i.e.~$\Theta^2=\Theta_\alpha^2$ for either $\alpha = e,\,\mu$ or $\tau$. 
In Fig.~\ref{fig:ScanLG_NH_dirac}, as long as large mixings are 
considered, i.e.~$|\xi|\gg 1$, we find that LG is compatible with the 
condition $\Theta^2_\tau> \Theta_\mu^2>\Theta_e^2$ (see further). 
For this reason, we only consider the bounds on $\Theta^2_\tau$ when showing 
the region excluded from past and present searches \cite{CHARM:1985anb, DELPHI:1996qcc, T2K:2019jwa, ArgoNeuT:2021clc, BaBar:2022cqj, Barouki:2022bkt} 
(upper grey region) and BBN \cite{Sabti:2020yrt, Boyarsky:2020dzc} (yellow). 
Moreover, we project the expected sensitivities on $\Theta^2_\tau$ 
of upcoming and proposed experiments \cite{FASER:2018eoc, Dib:2019tuj, Aielli:2019ivi, Batell:2020vqn, HIKE:2022qra, Abdullahi:2022jlv, Aberle:2839677, MATHUSLA:2020uve, Antel:2023hkf} 
(purple dot-dashed line). 
The prospective sensitivity on $\Theta^2$ of the 
discussed FCC-ee \cite{FCC:2018byv, FCC:2018evy} is also reported 
(green dashed line). }

\item{The maximal allowed values of $\Theta^2$ compatible with viable LG with 
LECPV from $\delta$ depend on the value of the Dirac phase.
For the case in Fig.~\ref{fig:ScanLG_NH_dirac} with $\delta = 3\pi/2$, 
these are $\Theta^2 \simeq 9\times 10^{-6},\, 5\times 10^{-7},\, 6\times 10^{-9},\,9\times 10^{-12}$ when $M_1 = 0.1, 1,\,10,\,100\,\text{GeV}$, respectively. 
By fixing $\delta = 195^\circ\, (345^\circ)$, we find $\Theta^2 \simeq 2\times 10^{-6}\,(1.5\times 10^{-5})$, $9\times 10^{-8}\,(1.5\times 10^{-6})$, $2\times 10^{-9}\,(1.2\times 10^{-8})$, $6\times 10^{-12}(4\times 10^{-11})$ at $M_1 = 0.1$, $1$, $10$, $100\,\text{GeV}$. 
We compare these results with the case of 
CICPV fixing $\omega = \pi/4$ or $3\pi/4$, $\delta = 3\pi/2$ and $\alpha_{23} = \pi$, so to maximise the CP-asymmetry and the maximal allowed mixing (see, e.g., \cite{Drewes:2016jae, Eijima:2018qke, Klaric:2021cpi}). 
We get $\Theta^2 \simeq 3\times 10^{-5}$, $3\times 10^{-6}$, 
$2.5\times 10^{-8}$, $4\times 10^{-11}$ at $M_1 = 0.1$, $1$, $10$, 
$100\,\text{GeV}$ (see also the results of 
\cite{Klaric:2020phc, Klaric:2021cpi} for comparison). We note, however, that the maximal allowed values of $\Theta^2$ in the case of CICPV do not exhibit strong dependence on $\delta$ and $\alpha_{23}$. The differences in the values 
obtained with LECPV from $\delta$ 
and from CICPV
reveal a separation between the parameter spaces of successful 
LG in the two cases.
The magnitude of this gap depends on $\delta$ and $M_1$.}

\item{We show in Fig.~\ref{fig:MxRatiosLG} the possible values of the mixing ratios 
$\Theta^2_\alpha/\Theta^2$ in a ternary plot. The four triangular regions 
in the plot are obtained for $\alpha_{23(21)} = \pi$ and $\omega = 0$, and
by marginalising over $\delta$ in the range $[0,\pi]$, 
(or, equivalently, $[\pi,2\pi]$), with the green and blue (yellow and red) 
triangles corresponding respectively to $\xi \geq0$ and $\xi \leq 0$ in the 
NO (IO) case. In such triangular regions, we find viable LG with LECPV from $\delta$. For $|\xi|\gg 1$, the triangles reduce to the shorter 
solid edges, while the intersection points correspond to $\xi = 0$. 
The larger and fainter blue (red) region, overlapping with the triangles associated to LECPV, is obtained by varying 
$\delta$, $\alpha_{23(21)}$ and
$\xi$ within their entire allowed ranges of 
possible values, and here LG is viable with additional sources of CP-violation from the Casas-Ibarra matrix and/or the Majorana phases. In the NO case, one has $\Theta_{\mu,\tau}^2>\Theta^2_e$, and, 
depending on whether $\xi \gg 1$, $\ll -1$ or $\sim 0$, either 
$\Theta^2_\mu>\Theta^2_\tau$, $\Theta^2_\tau>\Theta^2_\mu$ or 
$\Theta^2_\tau\sim \Theta^2_\mu$.}

\item{Concentrating on the NO case, we scan the LG space
 over $\delta$ across the entire ranges of 
masses and splittings considered. We find the results to be symmetric under the simultaneous change $\delta \to \delta \pm \pi$ and $\xi \to -\xi$. Moreover, the present $\eta_B$ 
can be reproduced with the correct sign only for i) $\xi> 0$ and 
$0 < \delta < \pi$, or ii) $\xi< 0$ and 
$\pi < \delta < 2\pi$. When large 
mixings are considered, i.e.~$|\xi|\gg 1$, the above two cases correspond 
respectively to \begin{enumerate}[i)] 
\item {$\Theta^2_\mu>\Theta^2_\tau> \Theta^2_e$ with $0.005\lesssim \Theta^2_e/\Theta^2\lesssim 0.12$, $0.69\lesssim \Theta^2_\mu/\Theta^2\lesssim 0.76$ and $0.19\lesssim \Theta^2_\tau/\Theta^2\lesssim 0.24$;} 
\item{$\Theta^2_\tau>\Theta^2_\mu> \Theta^2_e$ with $0.005\lesssim \Theta^2_e/\Theta^2 \lesssim 0.12$, $0.13\lesssim \Theta^2_\mu/\Theta^2\lesssim 0.16$ and $0.75\lesssim \Theta^2_\tau/\Theta^2 \lesssim 0.83$.}
\end{enumerate}

The situation is more involved for the IO spectrum: a shift of sign in $\xi$ changes prominently the mixings and the leading order CP-invariant in the electron flavour vanishes for 
LECPV from $\delta$. The IO case will be discussed in more details elsewhere.}
\end{itemize}
\begin{figure}
        \centering
        \includegraphics[width = 0.476\textwidth]{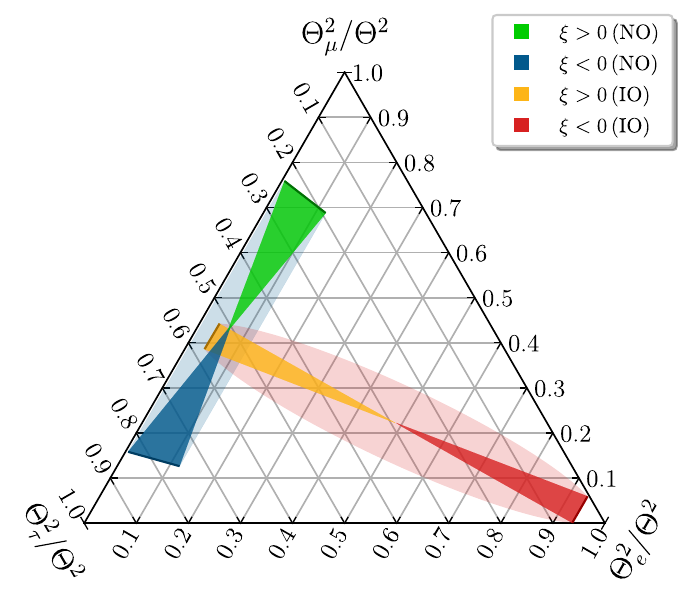}
        \caption{A ternary plot illustrating the ratios $\Theta_e^2/\Theta^2$ 
(lower axis), $\Theta_\mu^2/\Theta^2$ (right axis), and 
$\Theta_\tau^2/\Theta^2$ (left axis). The triangular regions correspond to: 
$\xi > 0$ and NO (green), $\xi < 0$ and NO (blue), $\xi > 0$ and IO (orange), 
and $\xi < 0$ and IO (red). The other parameters are: 
$\alpha_{23(21)} = \pi$ for the NO (IO) case and 
$\delta$ varied in the range $[0,\pi]$ (or, equivalently, $[\pi,2\pi]$). 
The fainter blue (red) region represents
the results obtained by varying $\xi$, $\delta$, $\alpha_{23(21)}$ over their entire ranges of allowed values (note that $\Theta_{e,\,\mu,\,\tau}^2 / \Theta^2$ do not depend on $\omega$ nor $M_{1,\,2}$).}
        \label{fig:MxRatiosLG}
    \end{figure}

The results were obtained for $\omega = 0$ and $\alpha_{23(21)}=\pi$. 
Everything would be the same for $\omega = \pi,\,2\pi$, while setting 
$\alpha_{23(21)} = 3\pi$ would imply equivalent results provided that 
the overall sign of $\xi$ is changed. An overall sign shift can be obtained by choosing $\omega = \pi/2, 3\pi/2$. \\

\noindent \textit{\textbf{Conclusions}}--- 
The results we have found indicate quite remarkably not only 
that LG with low-energy CP-violation solely from $\delta$ 
is viable in the mass range $0.1\leq M_1/\text{GeV}\leq 100$, but also that it 
is compatible with rather large values of $\Theta^2$. As the sensitivity 
reaches of proposed experiments enter inside the region of viable LG in 
the entire considered mass range, they could potentially probe the parameter 
space of the LG scenario discussed in this work. Moreover, we find 
viable LG for broad ranges of $\delta$ values within $0 < \delta < \pi$ and $\pi < \delta < 2\pi$. Qualitatively similar results hold in the IO case as well.

We have found a correspondence between the sign of the BAU and that 
of $\sin\delta$ in the NO case, which is reflected in the differences in the 
flavour hierarchies. More specifically, LG with LECPV 
from $\delta$ is successful in reproducing the positive BAU for either 
$0 < \delta < \pi$ ($\pi < \delta < 2\pi)$ and
$\Theta^2_\mu>\Theta^2_\tau > \Theta^2_e$ 
or $\pi < \delta < 2\pi$ ($0 < \delta < \pi$) and
$\Theta^2_\tau>\Theta^2_\mu > \Theta^2_e$ for $\omega = 0, \pi, 2\pi$ ($\pi/2, 3\pi/2$). As the physical observables 
at direct searches of heavy neutral leptons depend on the ratios 
$\Theta^2_\tau:\Theta^2_\mu:\Theta^2_e$, the
above
cases are phenomenologically 
different. Possible future signatures favouring a certain flavour hierarchy 
and a measurement of $\delta$ establishing  whether 
$0 < \delta < \pi$ or $\pi < \delta <  2\pi$
could  
discriminate between the
scenarios
considered in this work. Additionally, if experiments suggest a flavour structure outside the green and blue (yellow and red) triangles of Fig.~\ref{fig:MxRatiosLG} in the NO (IO) case, but still inside the light-blue (light-red) regions, LG would necessitate of additional sources of CP-violation, either from the Majorana phases and/or the Casas-Ibarra matrix.

Finally,
we have shown that there is a gap between the parameter spaces of LG 
with LECPV and CICPV, with the separation depending on $\delta$ and $M_1$. 
A measurement of $\delta$ and $\Theta^2$ at a certain mass scale 
in the associated gap would indicate the necessity of having additional 
sources of CP-violation other than $\delta$.

Overall, our results show that high-precision measurements of $\delta$, $\Theta^2$ and/or the ratios $\Theta_\tau^2:\Theta^2_\mu:\Theta^2_e$ will be crucial for understanding whether, within the scenario of LG we are considering, the Dirac CP-violating phase of the PMNS matrix can be the unique source of CP-violation, or additional sources coming from the Casas-Ibarra matrix and/or the Majorana phases are required in order to explain the presently observed BAU. \\

\noindent{\bf \textit{Acknowledgements}---}
 We thank S.~Sandner
 and B.~Shuve for useful email exchanges. A.G.~is grateful to the Kavli IPMU for the kind hospitality offered during the first part of this project. We acknowledge the use of computational resources from the parallel computing cluster of the Open Physics Hub (\href{https://site.unibo.it/openphysicshub/en}{https://site.unibo.it/openphysicshub/en}) at the Physics and Astronomy Department in Bologna. This work was supported in part by the European Union's Horizon research and innovation programme under the Marie Skłodowska-Curie grant agreements No.~860881-HIDDeN and No.~101086085-ASYMMETRY, and by the Italian INFN program on Theoretical Astroparticle Physics. S.T.P.~acknowledges partial support from the World Premier International Research Center Initiative (WPI Initiative, MEXT), Japan.
 

%

\end{document}